\documentclass[sn-nature, iicol]{sn-jnl}

\usepackage{graphicx}% 図用
\usepackage{amsmath}% 数式用
\usepackage{amsfonts}% 数式フォント用
\usepackage{comment}
\usepackage{color, xcolor}

\begin{document}

\title{Fiber-optic quantum interface with an array of more than 100 individually addressable atoms on an optical nanofiber}

\author[1]{\fnm{Mitsuyoshi} \sur{Takahata}}
\author[1]{\fnm{Jameesh} \sur{Keloth}}
\author[1]{\fnm{Takashi} \sur{Yamamoto}}
\author[1]{\fnm{Ken-ichi} \sur{Harada}}
\author[2]{\fnm{Shigehito} \sur{Miki}}
\author*[1,3]{\fnm{Takao} \sur{Aoki}}\email{takao@waseda.jp}

\affil[1]{\orgdiv{Department of Applied Physics}, \orgname{Waseda University}, \orgaddress{\street{3-4-1 Okubo}, \city{Shinjuku}, \state{Tokyo}, \postcode{169-8555}, \country{Japan}}}

\affil[2]{\orgdiv{Advanced ICT Research Institute}, \orgname{National Institute of Information and Communications Technology}, \orgaddress{\street{588-2 Iwaoka}, \city{Kobe}, \state{Hyogo}, \postcode{651-2492}, \country{Japan}}}

\affil[3]{\orgdiv{RIKEN Center for Quantum Computing (RQC)}, \orgname{RIKEN}, \orgaddress{\street{2-1 Hirosawa}, \city{Wako}, \state{Saitama}, \postcode{351-0198}, \country{Japan}}}

\abstract{
Integrating the scalability of individually addressable arrays of optical-tweezer-trapped single atoms with the efficient light-matter interface provided by nanophotonic waveguides has been a long-standing challenge in quantum technologies based on atoms and photons. 
Here we realize a quantum interface between photons guided in an optical nanofiber with a diameter of 310 nm and an array of on average 155 individually addressable atoms. 
Using a spatial light modulator and an objective lens with $NA = 0.45$, single cesium atoms are trapped in a one-dimensional array of 200 optical tweezer spots with micrometer-scale trap sizes on the nanofiber. 
Individual atoms are addressed by spatially scanning an excitation laser beam, focused to a spot size comparable to that of the traps through the same objective lens, along the nanofiber. 
We confirm the single-atom nature of the individual trapping sites through photon-correlation measurements of the guided fluorescence, observing strong photon antibunching with $g^{(2)}(0) \approx 0.26$.
We measure trap lifetimes of a few hundred milliseconds, with a maximum value of 460 ms, at an atom-surface separation of 670 nm without active cooling, representing an order-of-magnitude improvement over previous nanofiber traps. 
This platform opens a new regime for atom-photon interfaces, paving the way for scalable distributed quantum computing and quantum networks, as well as for the exploration of collective radiative effects in waveguide QED with individually addressable atoms.
}

\keywords{single atom, optical tweezer, optical waveguide}

\maketitle

% 実験セットアップとシステムの特性評価
\begin{figure*}[tb]
\centering
\includegraphics[width=\textwidth]{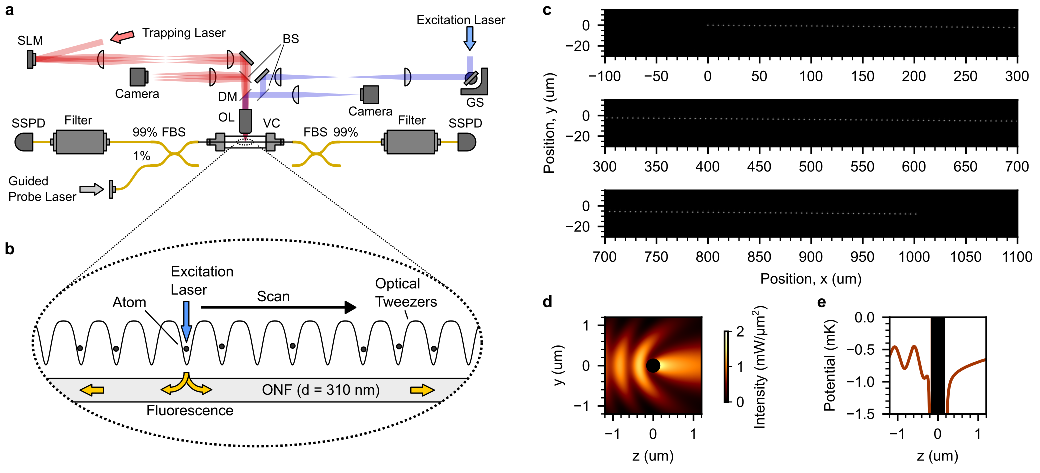}
\caption{
\textbf{Experimental setup for nanofiber-integrated optical tweezer array.}
\textbf{a,} Schematic of the optical arrangement. A spatial light modulator (SLM) generates the tweezer array, while a galvanometer scanner (GS) steers the excitation beam. Both are focused onto the optical nanofiber (ONF) via a high-NA objective lens (OL). The ONF is connected to a fiber detection setup with 99:1 fiber beam splitters (FBSs) and superconducting single-photon detectors (SSPDs).
\textbf{b,} Magnified view of the trapping geometry.  Fluorescence coupled into the guided mode is detected by the SSPDs.
\textbf{c,} Reflection image of the optical tweezer array (shown in segments). The 200 sites exhibit a uniform $5$-$\mu\text{m}$ pitch and high intensity uniformity.
\textbf{d,} Calculated distribution of the optical intensity, assuming a trapping laser ($\lambda = 935$ nm) focused to a beam waist of 1.0 $\mu$m. The trap laser forms a standing wave with its reflection from the ONF. 
\textbf{e,} Line profile of the trapping potential along the tweezer beam axis, given by the sum of the optical dipole potential and the attractive surface potential (see the Supplementary Note 1 for details).
}
\label{fig:setup}
\end{figure*}

Efficiently coupling many individually addressable single atoms to the guided modes of nanophotonic devices is a cornerstone for developing quantum technologies based on atoms and photons, such as distributed quantum computing and quantum networks, as well as for studying fundamental physics in waveguide quantum electrodynamics (QED) \cite{Cirac1999, Kimble2008, Monroe2014, Lodahl2015, Roy2017, Chang2018, Wehner2018, Awschalom2021, Lodahl2021, Sheremet2023, Azuma2023, Covey2023, Gonzalez-Tudela2024}. 
Cold atoms coupled to optical nanofibers offer a promising route toward this goal, providing strong atom-light interactions over extended lengths \cite{Tong2003, Sague2007, Brambilla2010, Tong2012, Solano2017a, Nayak2018, Li2024} and extremely efficient coupling to single-mode optical fibers through adiabatic tapers, with efficiencies exceeding 99.9\% \cite{Love1986, Love1991, Birks1992, Hoffman2014, Nagai2014, Ruddell2020}. 

In particular, two-color evanescent traps \cite{Kien2004} have been highly successful in this regard, capable of simultaneously confining thousands of atoms along the fiber waist \cite{Vetsch2010, Goban2012, Kestler2023}. This capability enabled demonstrations of, \textit{e.g.}, directional and controllable spontaneous emission of photons from trapped atoms into the nanofiber \cite{Mitsch2014}, storage and retrieval of light using electromagnetically induced transparency \cite{Sayrin2015, Gouraud2015}, the preparation of a single collective excitation in atomic arrays \cite{Corzo2019}, and cascaded interactions of guided light with ensembles of up to 1000 atoms \cite{Liedl2023}. 
However, a major limitation of these schemes is the lack of flexible control over the atomic geometry. They rely on fixed lattice constants determined by standing-wave potentials along the fiber axis, hindering the ability to arbitrarily define interatomic spacings or to deterministically address individual atoms.

In contrast, arrays of single atoms trapped in optical tweezers have emerged as a powerful platform offering individual addressability, scalability, reconfigurability, and high-fidelity control \cite{Endres2016, Barredo2016, Manetsch2025, Kusano2025}. This platform has enabled remarkable recent progress in neutral-atom quantum computing \cite{Ebadi2022, Bluvstein2022, Graham2022, Singh2023, Evered2023, Scholl2023, Ma2023, Bluvstein2024, Cao2024, Finkelstein2024, Grinkemeyer2025, Evered2025, Chiu2025, Bluvstein2026}, and provides a versatile system for exploring many-body physics \cite{Browaeys2020, Ebadi2021, Scholl2021, Spar2022}. 
Combining optical tweezer arrays with waveguide QED would therefore provide a highly scalable, controllable, and efficient atom-photon quantum interface \cite{Menon2024}.

Here, we demonstrate a fully integrated quantum interface consisting of a one-dimensional array of single cesium atoms trapped in optical tweezers in the near field of an optical nanofiber (Fig.~1). By combining holographic optical trapping with the efficient waveguide coupling of the nanofiber, we realize a system supporting 200 individually addressable atomic sites with an inter-site spacing of 5~$\mu$m and a distance of 670~nm from the nanofiber surface. 
This geometry ensures efficient coupling of the trapped atoms to the guided mode while maintaining the ability to spatially resolve and individually manipulate each atom, providing a scalable architecture for atom-light interaction. Using this platform, we successfully load on average 155 single cesium atoms into the tweezer array, achieving typical (maximum) trap lifetimes of a few hundred (460) milliseconds. 
We demonstrate site-resolved detection of atoms via the nanofiber guided mode and reconstruct the spatial distribution of atoms with high fidelity. This realization of a large-scale, individually addressable atom-waveguide interface opens new avenues for distributed quantum computation and quantum networks \cite{Djordjevic2021, Sunami2025, Sinclair2025, Main2025, Ji2026}, as well as for investigating collective effects with photon-mediated infinite-range interactions in waveguide QED \cite{Chang2012, Chang2013, Douglas2015, Gonzalez-Tudela2015, Corzo2016, Lodahl2017, Solano2017b, Bello2019, Albrecht2019, Mahmoodian2020, Fayard2021}.

\section*{Results}

Our experimental setup integrates a reconfigurable optical tweezer array with an optical nanofiber inside an ultra-high vacuum chamber, as illustrated in Fig. 1a and b. 
A linear array of 200 holographic optical tweezer traps\cite{Reicherter1999} is created with a laser at the magic wavelength of 935 nm for cesium D2 (6S$_{1/2}$ → 6P$_{3/2}$) transition\cite{McKeever2003} and a spatial light modulator (SLM).
These beams are focused onto the nanofiber using an objective lens with a numerical aperture of 0.45.
Figure 1c displays the reflection image of the tweezer beams off the nanofiber surface, revealing a periodic structure of 200 distinct sites with a uniform pitch of 5 $\mu$m. The intensity distribution across the array is highly uniform, confirming the precise holographic generation of the potentials. 
Single atoms are loaded into the tweezer trap array from cold cloud of atoms, prepared with magneto-optical trap (MOT) followed by gray molasses cooling (GMC) (see methods).
Crucially, the trapping mechanism relies on the interference between the incident tweezer beam and its reflection from the nanofiber surface (Fig. 1d and e). This interference creates a standing-wave potential that provides tight confinement in the radial direction\cite{Thompson2013, Nayak2019}, positioning the first trap minima that can stably trap atoms approximately 670 nm from the fiber surface for the current setting. This geometry ensures that the trapped atoms reside within the evanescent field of the nanofiber guided mode, providing high efficiency in atom-waveguide coupling\cite{Kien2005}. 
Note that, for the current trap beam powers, the trap minima at $\approx 190$ nm is not sufficiently deep to stably trap atoms due to the attractive potential from the surface (see the Supplementary Note 1 for details). Both ends of the fiber are directed to superconducting single-photon detectors (SSPDs)\cite{Miki2013}, enabling high-efficiency detection of fluorescence photons from atoms coupled into the nanofiber, or of the transmission and reflection of the probe beam launched into the fiber.

% 主要な結果
\begin{figure*}[tb]
\centering
\includegraphics[width=\textwidth]{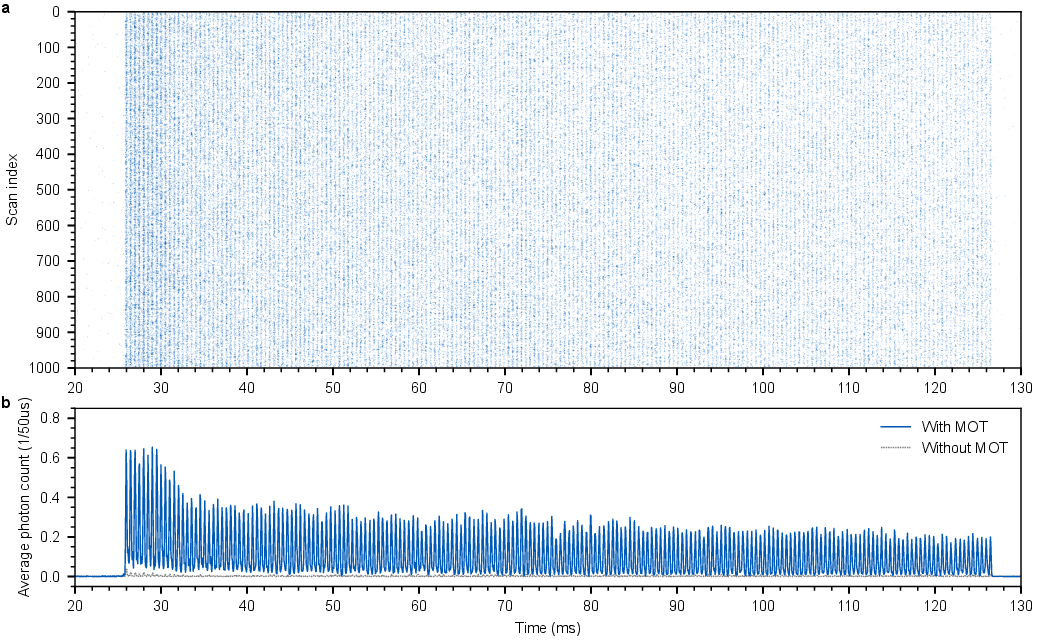}
\caption{
\textbf{Site-resolved fluorescence traces from atoms trapped in 200 tweezer array.}
\textbf{a,} 
Photon detection traces recorded while linearly scanning the excitation laser along the optical tweezer array. Each row corresponds to a single scan, and 1000 repetitions are shown. Blue dots indicate photon detection events within 50~$\mu$s time bins. The data show that single atoms trapped at individual sites can be clearly detected within a single scan. 
\textbf{b,} 
Averaged photon counts per time bin obtained from the 1000 traces shown in a (blue solid line), together with control measurements taken in the absence of the MOT (black dotted line). 
The observation of 200 distinct peaks, corresponding to the sequential addressing of each tweezer site, is consistent with atom trapping across all 200 sites and demonstrates the capability for individual addressing over the entire array.
}
\label{fluorescence}
\end{figure*}

First, we demonstrate the capability to address individual atoms within the large-scale array. We scan a focused resonant excitation laser beam across the chain of 200 sites using a galvanometer scanner, while simultaneously monitoring the photon counts coupled into the guided mode of the nanofiber with SSPDs. 
Figure~2a shows photon detection traces obtained during linear scanning of the excitation laser along the optical tweezer array over 1000 scans. Each row corresponds to a single scan, and the data demonstrate that single atoms trapped at individual sites are clearly resolved within a single scan. 
Figure~2b shows the photon counts averaged over these 1000 scans, which exhibit well-defined peaks corresponding to the individual trapping sites, thereby demonstrating the ability to excite and detect atoms across all 200 sites in a spatially resolved manner. 
The gradual reduction of the peak height is mainly attributed to the finite trap lifetime of a few hundred milliseconds, which is discussed below.

% 主要な結果の詳細な解析
\begin{figure}[tb]
\centering
\includegraphics[width=\columnwidth]{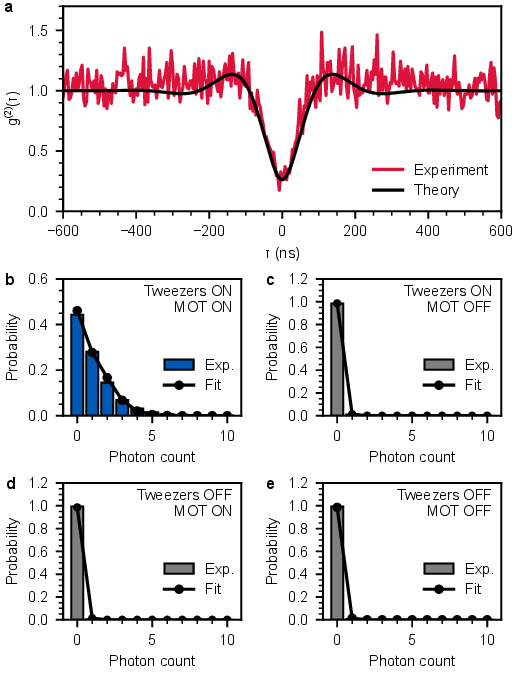}
\caption{
\textbf{Verification of single-atom trapping and estimation of filling factor.}
\textbf{a,} Second-order photon correlation function, $g^{(2)}(\tau)$, integrated over the array and 31000 scans. The blue line represents the experimental results, and the black line represents the theoretical calculation (see Supplementary Note 2). The observed antibunching $g^{(2)}(0) < 0.5$ confirms the single-atom nature of each trapped emitter.
\textbf{b-e,} Histograms of photon counts recorded for individual trap sites. For \textbf{b}, the data aggregates the counts obtained from all 200 sites across 31000 measurement repetitions, while \textbf{c-e} are from 1000 repetitions. The panels correspond to the following experimental conditions: with optical tweezers and MOT (\textbf{b}); with optical tweezers but without MOT (\textbf{c}); without optical tweezers but with MOT (\textbf{d}); and without optical tweezers or MOT (\textbf{e}).
The clear separation of the signal distribution in b from the control measurements (\textbf{c-e}) allows for the determination of the filling factor (loading efficiency), yielding a total estimated atom number of $N \approx 155$.
}
\label{fluorescence}
\end{figure}

To definitively verify the single-atom nature of the trapped emitters, we measure the second-order photon correlation function, $g^{(2)}(\tau)$, of the guided fluorescence \cite{Kimble1977, Short1983} for the above measurements. The correlation statistics are integrated over the entire array for 31,000 scans. The resulting data (Fig.~4a), obtained by normalizing the overall intensity envelope (see Supplementary Note~2), is in excellent agreement with the theoretical calculation based on the experimental parameters. Most importantly, we observe a strong suppression of coincidence events at zero time delay, yielding a value of $g^{(2)}(0) \approx 0.26$. This pronounced antibunching provides clear evidence that the optical tweezers contain single atoms rather than ensembles.

Having confirmed the single-atom nature of the fluorescence signal, we next analyze the loading statistics across the entire array to determine the number of trapped atoms. Photon-count histograms are recorded for each site under four experimental conditions: with and without optical tweezers, and with and without the cold atom cloud (Fig.~4b--e). Due to the limited photon counts per site, a clear bimodal distribution separating the zero- and one-atom manifolds \cite{Schlosser2001} is not fully resolved. 
Nevertheless, as quantified by the control measurements shown in Fig.~4c--e, the background level is extremely low, resulting in a high signal-to-background ratio. 
This enables a reliable estimate of the loading probability by fitting the histograms with Poissonian distributions. Specifically, for the control measurements, we obtain mean values of 0.0137, 0.0054, and 0.0056 for Fig.~4c, d, and e, respectively. We then fit the data with atoms (Fig.~4b) using a sum of two Poissonian distributions,
\[
P(X=k)= w \frac{\mu_a^k e^{-\mu_a}}{k!} + (1-w) \frac{\mu_b^k e^{-\mu_b}}{k!},
\]
where we set $\mu_b = 0.0137$ obtained above, and $w$ and $\mu_a$ are free parameters. From the fit, we obtain $\mu_a = 1.22$ and $w = 0.775$. Using the extracted filling factor of 0.775, we estimate an average number of $N = w \times 200 = 155$ trapped atoms.

Furthermore, we derive a strict lower bound on the number of trapped atoms. Specifically, we attribute the reduction in the fraction of sites with zero detected photons (from 98.5\% in the control condition to 44.3\% in the loading condition) to the presence of trapped atoms, yielding $N_{\rm lower-bound} = 108$. 
Because there remains a finite probability of detecting zero photons even from an occupied site, while the probability of detecting non-zero photons from an empty site is only 0.00137, this procedure provides a conservative estimate of the filling factor. This analysis therefore establishes that more than 100 atoms are trapped in the array.

We next estimate the atom-waveguide coupling efficiency $\beta$ from transmission measurements. Figure~4a shows the optical depth spectrum of the nanofiber with single atoms loaded into the tweezer traps, measured by scanning the frequency of a weak probe laser across the $F=4 \rightarrow F^\prime = 5$ transition. A clear peak is observed at the free-space atomic resonance frequency, confirming trapping at a magic wavelength with negligible differential AC Stark shifts. 

From the observed peak optical depth of $OD = -\ln(T) = 1.2$ and the total atom number $N = 155$ derived from the fluorescence histograms, we obtain an average single-atom optical depth of $d_0 = OD/N \approx 0.077$. In the weak-excitation limit, this corresponds to an effective coupling efficiency of $\beta = d_0/2 \approx 0.38\%$. This experimentally derived value is in reasonable agreement with our theoretical estimate for atoms confined in the secondary potential minimum ($\beta \approx 0.6\%$, see Supplementary Note~1).

% システムのロバスト性の検証
\begin{figure}[tb]
\centering
\includegraphics[width=\columnwidth]{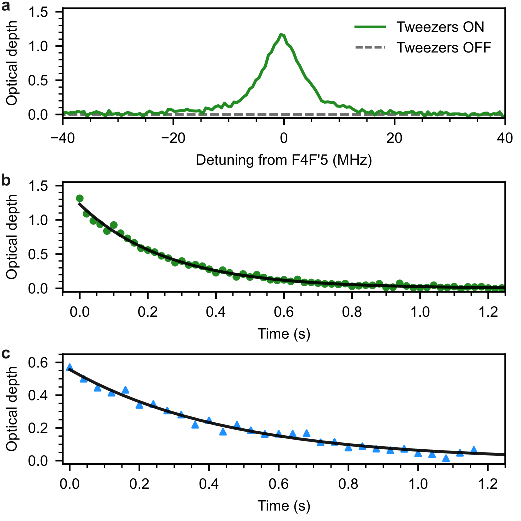}
\caption{
\textbf{Transmission measurements through the nanofiber.}
\textbf{a,} Optical depth spectrum measured via the nanofiber guided mode. The peak at the atomic resonance frequency in free space verifies the magic-wavelength operation of the 935-nm trapping light.
\textbf{b,} Time evolution of the optical depth (OD) at the atomic resonance. The OD is directly proportional to the number of trapped atoms, $N$, serving as a probe for the trap lifetime. The data points represent the measured OD at the atomic resonance after a variable hold time in the optical tweezers with cooling beams extinguished. The solid line indicates a single-exponential fit to the data, yielding a $1/e$ lifetime of $\tau = 0.26~\text{s}$. This duration significantly exceeds the lifetimes typically achieved in conventional nanofiber-based trapping schemes, demonstrating the robust stability of the integrated platform.
\textbf{c,} The observed results with the maximum value for the lifetime, $\tau = 0.46~\text{s}$, with 100 tweezer sites for another nanofiber with a diameter of 280 nm.
}
\label{fluorescence}
\end{figure}

Finally, we evaluate the stability of the trapped atoms by measuring their lifetime in the optical tweezers. The observed optical depth, which is proportional to the number of trapped atoms, decays exponentially with a characteristic $1/e$ lifetime of $\tau = 0.26~\text{s}$ (Fig.~4b). Notably, this lifetime is significantly longer than those typically reported for atoms trapped in the evanescent fields of nanofibers without continuous cooling, which are generally limited to several tens of milliseconds \cite{Vetsch2010, Meng2018, Corzo2019}. 

We reduce the background cesium vapor pressure but observe no significant improvement in the lifetime, suggesting that background gas collisions are not the dominant loss mechanism. 
Measurements performed under varying experimental conditions yield lifetimes that differ by a factor of order unity, yet consistently remain on the scale of a few hundred milliseconds. 
In particular, we observe a maximum lifetime of $\tau = 0.46$~s with 100 tweezer sites for another nanofiber with a diameter of 280 nm, as shown in Fig~4c. Further investigation of the mechanisms limiting the trap lifetime is left for future work.

\section*{Discussion}

Our demonstration of a large-scale, individually addressable atom-waveguide interface bridges the gap between two powerful technologies: optical tweezer arrays and waveguide quantum electrodynamics (QED). While free-space optical tweezers have revolutionized the deterministic and reconfigurable assembly of atomic arrays, integrating them with efficient fiber-based interconnects has remained a challenge. Conversely, previous nanofiber traps lacked the ability to arbitrarily define trap positions or address individual atoms due to sub-wavelength site spacings. Our platform overcomes this trade-off, offering both the scalability of waveguide systems and the microscopic control of optical tweezers.

We have successfully trapped 155 individually addressable atoms in the evanescent field of the nanofiber with a coupling efficiency of 0.38\%. Both the number of atoms and the coupling efficiency are currently limited by the available trapping-light power in our setup. By increasing the total power by a factor of five while keeping the power per tweezer beam constant, up to 1000 trap sites could be generated, assuming a 3-mm field of view for the objective lens and an inter-site spacing of 3 $\mu$m. On the other hand, increasing the power per tweezer beam would allow the coupling efficiency to be enhanced to 5.26\% (see Supplementary Note 1). Furthermore, it is straightforward to form a high-finesse inline cavity by inscribing a pair of fiber Bragg grating mirrors at both ends of the nanofiber\cite{Ruddell2020}. With such a cavity, the coupling efficiency could be boosted close to unity via Purcell enhancement.

The efficient light-matter interaction combined with site-selective control enables the realization of fiber-integrated multiplexed quantum memories and quantum processing units. This work thus establishes a foundational architecture for distributed quantum computing and long-distance quantum communication networks, where the efficient interface between stationary qubits and fiber-guided flying photons is paramount\cite{Kimble2008}. By providing a robust hardware platform that combines high-fidelity control with efficient optical interfacing, our result marks a key step toward scalable quantum interconnects\cite{Djordjevic2021, Sunami2025, Sinclair2025, Main2025, Ji2026}.

This platform also opens new avenues for exploring many-body physics in waveguide QED\cite{Chang2012, Chang2013, Douglas2015, Gonzalez-Tudela2015, Corzo2016, Lodahl2017, Solano2017b, Bello2019, Albrecht2019, Mahmoodian2020, Fayard2021}. 
With over 100 atoms strongly coupled to a single waveguide mode, the system is ideally poised to investigate collective radiative effects, such as Dicke superradiance and subradiance\cite{Dicke1954}, where the precise geometry of the array plays a critical role \cite{Solano2017b, Albrecht2019}. Near-unity Bragg reflection from atomic array with a period perfectly commensurate with the atomic resonant wavelength can be realized, in contrast to the reflectance up to 75\% that has been achieved with close-to-commensurate array with conventional guided traps\cite{Corzo2016}. This will enable $e.g.,$ the realization of the settings of cavity QED with atomic mirrors\cite{Chang2012}.

\section*{Methods}

\paragraph{Nanofiber fabrication.}
The optical nanofiber was fabricated by tapering a commercial single-mode fiber (Fibercore, SM800) using a heat-and-pull technique with a hydrogen-oxygen flame \cite{Ruddell2020}. The taper profile was designed to have a uniform waist diameter of 310~nm and a length of 5~mm. This waist diameter was selected to maximize the evanescent field coupling between the trapped cesium (Cs) atoms and the fundamental guided mode ($\text{HE}_{11}$), while the extended waist length accommodates a sufficiently large number of addressable trap sites. The entire fabrication process was optimized to ensure high optical transmission and mechanical robustness.

\paragraph{Atom preparation and cooling.}
The fabricated nanofiber was mounted inside an ultra-high vacuum (UHV) chamber maintained at a pressure of $\sim 10^{-8}$~Pa and filled with Cs vapor. To prepare the atomic ensemble, we employed a vapor-cell magneto-optical trap (MOT). An elongated MOT configuration was utilized to generate a cigar-shaped atomic cloud with a length of approximately 3~mm along the nanofiber axis, ensuring optimal spatial overlap with the nanofiber waist. Following the MOT phase, the atoms were further cooled using gray molasses cooling (GMC). The GMC beams had a total power of 20~mW and were blue-detuned by 30~MHz from the Cs $6S_{1/2}(F=4) \to 6P_{3/2}(F'=4)$ transition. A repumping laser tuned to the $F=3 \to F'=4$ transition was superimposed to prevent population accumulation in the dark state.

\paragraph{Generation of the optical tweezer array.}
The one-dimensional array of 200 individual atom trap sites was generated using a high-intensity, continuous-wave laser operating at a magic wavelength of 935~nm (FL-SF-935-10-CW, Precilaser) for the cesium D2 line. The wavefront of the trapping laser was phase-modulated by a spatial light modulator (SLM) (GAEA-2, Holoeye) to create the desired multi-trap pattern. The phase hologram was calculated using the Weighted Gerchberg-Saxton (WGS) algorithm \cite{Leonardo2007} to ensure uniform trap depths across the array. The modulated beam was focused onto the nanofiber surface inside the vacuum chamber using a high-numerical-aperture objective lens ($\text{NA} = 0.45$) tailored for the 6-mm thick vacuum window to ensure diffraction-limited performance. The polarization of the trapping laser was set parallel to the nanofiber axis. Finite-difference time-domain (FDTD) simulations confirmed that the interference between the incident beam and its reflection from the fiber surface creates stable standing-wave potential minima located approximately 670~nm from the nanofiber surface (see Supplementary Note 1). To load the atoms into the tweezers, the cold atomic ensemble was subjected to GMC for 100~ms, allowing the atoms to spatially overlap with the potential minima of the array. Subsequently, all cooling and repumping lasers were extinguished, and the system was held in the dark for at least 20~ms to allow untrapped or weakly trapped atoms to fall away from the trapping region under gravity.

\paragraph{Site-selective addressing and detection.}
For site-selective interrogation, we employed a focused external excitation laser comprising both pumping and repumping fields, resonant with the Cs $F=4 \to F'=5$ and $F=3 \to F'=4$ transitions, respectively, which was spatially scanned along the array using a galvanometer mirror and combined with the optical tweezer path via a dichroic mirror.

The power of the pumping field was set to achieve the saturation intensity at the position of the atoms.
To prevent reabsorption of fluorescence by other trapped atoms, all the atoms are initialized in the $F=3$ ground state by turning off the repumping light 5~ms before the cooling light at the end of the GMC stage.
Then the excitation spot was scanned along the nanofiber at a velocity of approximately 0.01~m/s, reaching the first atom of the array approximately 26~ms after the termination of the GMC stage. The polarization of the excitation beam was aligned parallel to the nanofiber axis to maximize the excitation efficiency. Crucially, the nanofiber does not propagate any guided trapping fields (neither attractive nor repulsive) during this process. It serves exclusively as a passive fluorescence collection probe, ensuring that the atoms are free from strong light shifts or heating effects typically caused by high-power guided light. When an atom trapped in a specific site was illuminated by the scanning beam, the fluorescence photons coupled into the guided mode of the nanofiber were guided out of the vacuum chamber. To isolate the fluorescence signal from the stray background light, these photons were spectrally filtered using a combination of a volume Bragg grating (VBG; BP-852-99, OptiGrate) and two bandpass filters (FBH852-3, FBH850-10, Thorlabs) before detection by a superconducting single-photon detector (SSPD). The SSPD converted the filtered photons into electrical pulses, and data acquisition was performed by time-tagging these pulses using a multi-scaler instrument (MCS8A, FAST ComTec) to reconstruct the fluorescence spatial profile.

\paragraph{Optical transmission measurements.}
We performed transmission measurements through the nanofiber to measure the optical depth and the trap lifetime. Following the GMC phase, a repumping field resonant with the $F=3 \to F'=4$ transition was applied for 5~ms to initialize the atomic population in the $F=4$ ground state. A weak probe laser (0.1~pW) was launched into one end of the nanofiber. The polarization of the probe field at the nanofiber waist was adjusted to be parallel to the radial axis connecting the nanofiber center and the trapped atoms to maximize the interaction cross-section. During these measurements, no repumping field was co-propagated to ensure that its presence did not affect the decay rate of the optical depth. The transmitted light collected at the other end of the nanofiber was guided to the SSPD through the same filtering stage used for the fluorescence measurements. For the transmission spectrum, the frequency of the probe laser was scanned across the Cs $F=4 \to F'=5$ resonance to resolve the absorption profile. For the lifetime measurement, the probe laser frequency was frequency-stabilized to the resonance peak, and the transmission signal was recorded as a function of time to monitor the atomic loss. In both measurements, data acquisition was triggered after a dark interval of $>20$~ms following the GMC phase to eliminate contributions from untrapped background atoms. The transmission was determined by normalizing the photon counts obtained with the trapped atoms against the reference counts measured without the atomic cloud.

\section*{Supplementary Note 1: Tweezer trap potentials and atomic coupling efficiencies}

The optical tweezer trap in our setup is formed by interference between the incident trapping beam and its reflection from the nanofiber surface. This interference creates a standing-wave potential that provides tight radial confinement near the nanofiber surface\cite{Thompson2013, Nayak2019}.

Figure 5a shows the calculated distribution of the trapping beam intensity in the plane perpendicular to the nanofiber axis, obtained using the finite-difference time-domain (FDTD) method. The first and second lattice sites (optical intensity maxima) are located at 190 nm and 671 nm from the nanofiber surface, respectively.

The total trapping potential is given by the sum of the optical dipole potential and the attractive van der Waals (vdW) potential near the fiber surface. The first lattice site is strongly influenced by the vdW potential, which significantly reduces the effective trap depth to $V_1$, as shown in Fig. 5b. For sufficiently low trapping-beam power, $V_1$ can even become negative, completely preventing stable trapping. In contrast, the vdW contribution at the second lattice site is negligible, and the corresponding trap depth $V_2$ is essentially determined by the optical potential alone. 
Figure 5c shows the calculated spatial distribution of the coupling efficiency $\beta$ for the $F=4 \leftrightarrow F^\prime = 5$ Cs D2 line, and Fig. 5d presents the line profile of $\beta$ along the radial ($z$) direction. At the potential minima, the calculated coupling efficiencies are 5.3\% for the first lattice site and 0.6\% for the second lattice site.

\begin{figure}[htbp]
\centering
\includegraphics[width=0.45\textwidth]{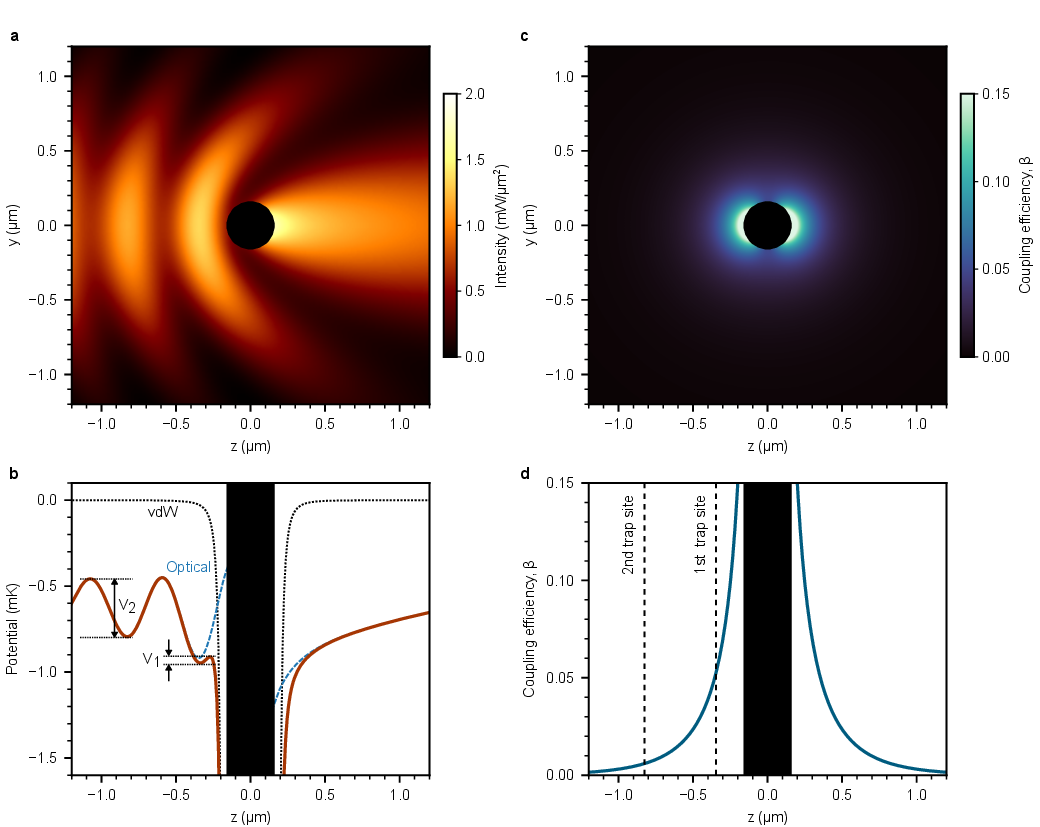}
\caption{
\textbf{Numerical analysis of the optical trapping potential and coupling efficiency.}
In all panels, the black filled areas (circles in a, b and shaded regions in c, d) indicate the optical nanofiber (diameter: 310 nm).
\textbf{a,} Calculated distribution of the optical intensity, assuming a trapping laser ($\lambda = 935$ nm) focused to a beam waist of 1.0 $\mu$m. 
\textbf{b,} Line profiles of the optical potential (red solid line), vdW potential (black dotted line), and the total potential (blue dashed line). 
For the optical potential, we assume the trap beam power of $P = 1.5$ mW. For the vdW potential, we use the $C_3$ coefficient corresponding to a Cs metallic surface\cite{Derevianko1999}, taking into account the adsorption of Cs atoms on the nanofiber surface. 
The first minimum of the optical potential at 190 nm from the surface is strongly affected by the vdW potential, resulting in the reduction of the effective trap depth to $V_1$, where as that for the second minimum at 671 nm from the surface is negligible.
\textbf{c,} Calculated distribution of the coupling efficiency $\beta$ between the Cs atom and the nanofiber.
\textbf{d,} Line profile of the coupling efficiency along the $z$-axis. The dashed lines indicate the position of the first and second trap sites.
}
\end{figure}

Atoms are loaded into the tweezer traps from a millimeter-scale cold-atom cloud. For low trap-beam powers, it is naturally expected that atoms are not efficiently loaded into the first lattice sites because the trap depth $V_1$ becomes insufficient to support stable confinement. In contrast, the second and higher-order lattice sites are expected to be populated with relatively uniform probabilities, because their trap depths are largely unaffected by the van der Waals potential, and because the spatial extent of the MOT cloud is much larger than the standing-wave period, so that atoms are loaded into successive lattice sites with nearly equal probability. However, since the coupling efficiencies for the third and higher-order sites are extremely small, their contribution to the detected signal is negligible. Consequently, at low trap-beam powers the signal originates predominantly from atoms in the second lattice sites, whereas at higher powers the contribution from the first lattice sites becomes significant.

\begin{figure}[p]
\centering
\includegraphics[width=0.45\textwidth]{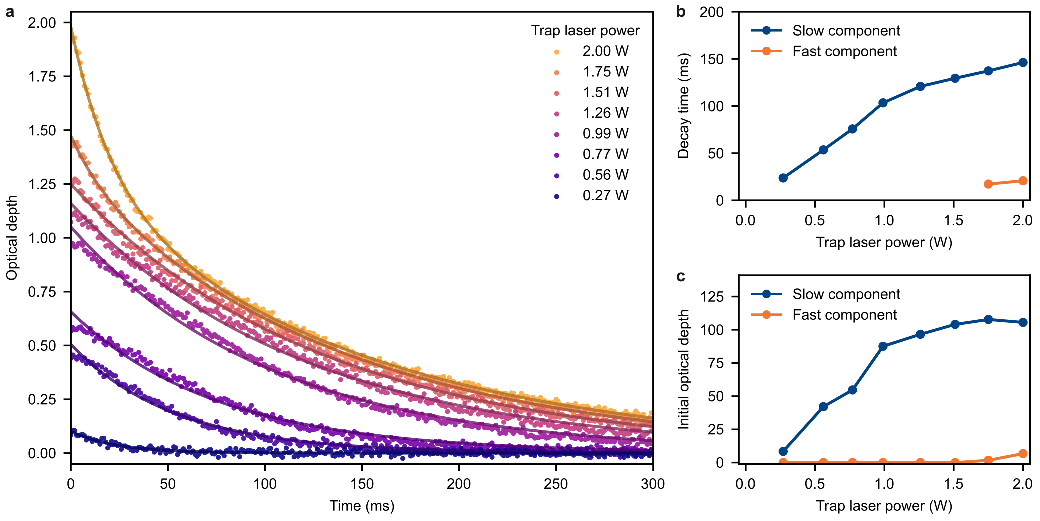}
\caption{
\textbf{Temporal traces of optical depth at the atomic resonance for various trap beam powers.}
\textbf{a,} Measured optical depth (OD) as a function of time for various optical tweezer powers, ranging from 0.27 W to 2.00 W. \textbf{b,} Extracted decay times and \textbf{c,} initial optical depths obtained by fitting the temporal traces in \textbf{a} with Equation (1). In both panels, the data are decomposed into a fast component ($\tau_1$ and $OD_1 = 2 \beta_1 N_1$, shown in orange) and a slow component ($\tau_2$ and $OD_2 = 2 \beta_2 N_2$, shown in blue).
}
\end{figure}

To investigate this transition, we measured the transmission of a weak resonant probe beam (with a power of 0.1~pW) through the nanofiber for various trap-beam powers $P$ per individual tweezer spot. Approximately 0.1\% of the total laser power was distributed to each spot, resulting in a range of $P$ from 0.27~mW to 2.00~mW. The measured temporal traces of the optical depth (OD) are shown in Fig.~6a. While the OD decay is well described by a single exponential for low trap powers, a clear double-exponential behavior appears as the power increases.

We attribute the long-lived component, present at all trap powers, to atoms in the second lattice sites. The short-lived component, observed only at higher trap powers, is attributed to atoms trapped in the first lattice sites. The atoms in the first sites are exposed to higher intensities of the probe beams and thus have shorter decay times.
Accordingly, we model the OD dynamics as
\begin{equation}
OD(t) = 2 \beta_1 N_1 e^{-t/\tau_1} + 2 \beta_2 N_2 e^{-t/\tau_2},
\end{equation}
where $N_{1(2)}$ denotes the initial number of atoms in the first (second) lattice sites, $\beta_{1(2)}$ is the corresponding coupling efficiency, and $\tau_{1(2)}$ is the decay time under continuous probing.

Using the theoretical values $(\beta_1, \beta_2) = (0.053, 0.006)$, we extracted the decay times $\tau_{1(2)}$ and the initial atom numbers $N_{1(2)}$ from fits to the measured OD curves, as summarized in Figs.~6b and 6c, respectively (see also Table~1). For trap powers $P \leq 1.5$~mW, we find that no atoms occupy the first lattice sites. For $P \geq 1.75$~mW, atoms begin to populate the first sites, and the occupancy increases with power. However, due to the limited available trap-beam power in the present setup, only approximately 4\% of the first lattice sites can be occupied. Therefore, all measurements presented in the main text were performed under conditions where atoms occupy only the second (and higher) lattice sites.

\begin{table}[htbp]
\centering
\begin{tabular}{|l|r|r|r|r|}
\hline
P (mW) & $N_1$ & $\tau_1$ (ms) & $N_2$ & $\tau_2$ (ms) \\
\hline
$0.27$ & 0 & -- & 8 & 24\\
$0.56$ & 0 & -- & 42 & 54\\ 
$0.77$ & 0 & -- & 55 & 76\\
$0.99$ & 0 & -- & 88 & 103\\
$1.26$ & 0 & -- & 97 & 121\\
$1.51$ & 0 & -- & 104 & 129\\
$1.75$ & 2 & 17 & 108 & 137\\
$2.00$ & 7 & 21 & 106 & 146\\
\hline
\end{tabular}
\caption{Extracted atom numbers and decay time constants for different trap-beam powers.
Fitted parameters $N_1$, $N_2$, $\tau_1$, and $\tau_2$ obtained from double-exponential fits to the measured optical-depth dynamics [Eq.~(1)] for various trap-beam powers $P$ (per individual tweezer spot). Here, $N_{1(2)}$ denotes the initial number of atoms in the first (second) lattice sites, and $\tau_{1(2)}$ represents the corresponding decay time under continuous probing. The coupling efficiencies $(\beta_1, \beta_2) = (0.053, 0.006)$ were fixed to their theoretically calculated values during the fitting procedure.}
\end{table}

\clearpage

\section*{Supplementary Note 2: Analysis of the second-order photon correlation function}

In this section, we describe the derivation and analysis of the second-order photon correlation function $g^{(2)}(\tau)$ shown in Fig.~4a of the main text. The analysis consists of two steps: (i) normalization of the raw coincidence histogram to remove macroscopic intensity modulation and (ii) comparison with a parameter-free theoretical model.

A continuous-wave probe beam is focused onto the nanofiber surface and scanned along the fiber axis at a speed of 0.01~m/s. Fluorescence photons are collected by superconducting single-photon detectors (SSPDs) at both ends of the nanofiber and recorded using a multiple-event time digitizer (MCS8A, FAST ComTec) with 0.8~ns time resolution.

Figure~7 shows the same coincidence dataset displayed at three different time scales.
On a 100~$\mu$s scale (Fig.~7a), a broad envelope is observed. Given the probe-beam spot size ($\sim 2$~$\mu$m) and the scan velocity, the geometric transit time across a trapped atom is approximately 200~$\mu$s, consistent with the observed envelope.
On a 10~$\mu$s scale (Fig.~7b), the correlation peak exhibits a width of approximately 4~$\mu$s, which is significantly shorter than the geometric transit time. This shortening is attributed to the heating induced by the near-saturation probe intensity, which rapidly heats and ejects the atom from the trap.
On a 1~$\mu$s scale (Fig.~7c), the nanosecond-scale quantum correlations near $\tau = 0$ are clearly resolved, while the macroscopic envelope remains approximately constant within the reference windows (500~ns $< |\tau| <$ 800~ns).

To extract the intrinsic photon statistics, we applied a local normalization procedure. The reference count level
\begin{equation}
\bar{C}{\mathrm{ref}} = \langle G^{(2)}(\tau) \rangle , 500 \mathrm{ns} < |\tau| < 800 \mathrm{ns}
\end{equation}
was obtained by averaging counts in these windows, where atomic correlations are negligible. The normalized correlation function is then given by
\begin{equation}
g^{(2)}(\tau) = \frac{G^{(2)}(\tau)}{\bar{C}_{\mathrm{ref}}}.
\end{equation}
This procedure removes the slow intensity envelope and reveals the antibunching dip and Rabi oscillations around $\tau = 0$.

The solid curve in Fig.~4a of the main text is not a fit but a theoretical prediction based solely on independently measured experimental parameters. For a resonantly driven two-level atom, the ideal second-order correlation function is\cite{WallsMilburn}
\begin{equation}
g^{(2)}_{\mathrm{ideal}}(\tau) = 1 - e^{-3\gamma |\tau| / 4} \left( \cos(\kappa |\tau|) + \frac{3\gamma}{4\kappa} \sin(\kappa |\tau|) \right),
\end{equation}
where $\gamma = 2\pi \times 2.61$~MHz is the transverse decay rate of the Cs D2 transition and
\begin{equation}
\kappa = \sqrt{(2\Omega)^2 - (\gamma/4)^2}.
\end{equation}
The Rabi frequency $\Omega$ was calculated assuming on-resonance excitation at approximately the saturation intensity ($s \approx 1$). The value $s \approx 1$ was independently verified experimentally from the excitation-intensity dependence of the fluorescence photon count rate. Under this condition, we obtain $\Omega = 2\pi \times 1.85$~MHz and $\kappa = 2\pi \times 3.64$~MHz.

To account for background counts (e.g., from the scattering of trap laser and the fluorescence of atoms in the third and higher lattice sites) that reduce the antibunching visibility, we introduce a small offset parameter $\delta$:
\begin{equation}
g^{(2)}(\tau) = (1 - \delta) g^{(2)}_{\mathrm{ideal}}(\tau) + \delta.
\end{equation}

The parameter $\delta \approx 0.26$ was determined from the experimentally observed value of $g^{(2)}(0)$. The excellent agreement between this parameter-free calculation (apart from the independently determined background fraction) and the experimental data provides strong evidence that a single atom is trapped at each occupied site, confirming the single-atom nature of the emitters in our system.

\begin{figure}[htbp]
\centering
\includegraphics[width=0.45\textwidth]{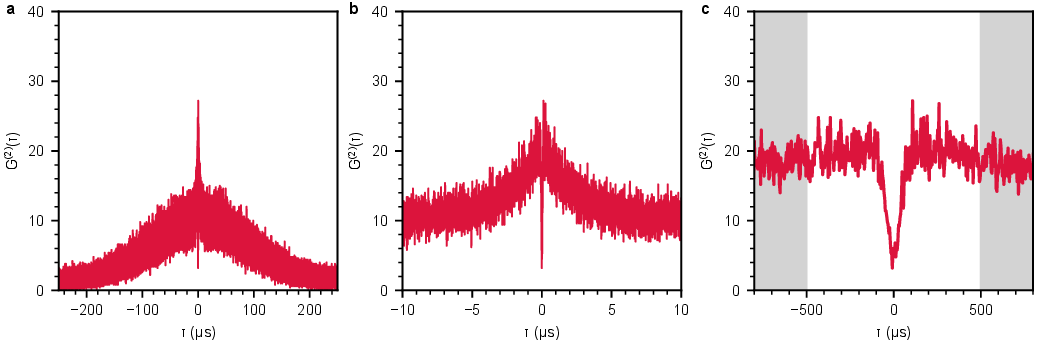}
\caption{
\textbf{Raw coincidence histograms of fluorescence photons at different time scales.}
The histograms display the same dataset with different time axis magnifications. 
\textbf{a,} 100 $\mu$s scale. The broad envelope corresponds to the geometric transit time ($\sim 200$ $\mu$s) of the scanning probe beam. 
\textbf{b,} 10 $\mu$s scale. The central peak has a width of $\sim 4$ $\mu$s, indicating the effective interaction time limited by recoil heating and trap loss. 
\textbf{c,} 1 $\mu$s scale. This view highlights the nanosecond-scale quantum correlations around $\tau=0$. The gray shaded areas ($500 < |\tau| < 800$ ns) indicate the reference windows used for normalization, where the macroscopic intensity is quasi-constant.
}
\end{figure}

\clearpage

%\bibliography{export20251226}% 参照文献ファイル（.bib）の名前

\begin{thebibliography}{99}

\bibitem{Cirac1999}
Cirac, J.~I. et al. Distributed quantum computation over noisy channels.
\emph{Phys. Rev. A} 59, 4249 (1999).

\bibitem{Kimble2008}
Kimble, H.~J. The quantum internet.
\emph{Nature} 453, 1023--1030 (2008).

\bibitem{Monroe2014}
Monroe, C. et al. Large-scale modular quantum-computer architecture with atomic memory and photonic interconnects.
\emph{Phys. Rev. A} 89, 022317 (2014).

\bibitem{Lodahl2015}
Lodahl, P. et al. Interfacing single photons and single quantum dots with photonic nanostructures.
\emph{Reviews of Modern Physics} 87, 347--400 (2015).

\bibitem{Roy2017}
Roy, D. et al. Colloquium: Strongly interacting photons in one-dimensional continuum.
\emph{Reviews of Modern Physics} 89, 021001 (2017).

\bibitem{Chang2018}
Chang, D.~E. et al. Colloquium: Quantum matter built from nanoscopic lattices of atoms and photons.
\emph{Reviews of Modern Physics} 90, 031002 (2018).

\bibitem{Wehner2018}
Wehner, S. et al. Quantum internet: A vision for the road ahead.
\emph{Science} 362, 303 (2018).

\bibitem{Awschalom2021}
Awschalom, D. et al. Development of Quantum Interconnects (QuICs) for Next-Generation Information Technologies.
\emph{PRX Quantum} 2, 017002 (2021).

\bibitem{Lodahl2021}
Lodahl, P. et al. Quantum-dot-based deterministic photon-emitter interfaces for scalable photonic quantum technology.
\emph{Nature Nanotechnology} 16, 1308--1317 (2021).

\bibitem{Sheremet2023}
Sheremet, A.~S. et al. Waveguide quantum electrodynamics: Collective radiance and photon-photon correlations.
\emph{Reviews of Modern Physics} 95, 015002 (2023).

\bibitem{Azuma2023}
Azuma, K. et al. Quantum repeaters: From quantum networks to the quantum internet.
\emph{Reviews of Modern Physics} 95, 045006 (2023).

\bibitem{Covey2023}
Covey, J.~P. et al. Quantum networks with neutral atom processing nodes.
\emph{Rpj Quantum Inf.} 9, 1 (2023).

\bibitem{Gonzalez-Tudela2024}
Gonzalez-Tudela, A. et al. Light-matter interactions in quantum nanophotonic devices.
\emph{Nature Reviews Physics} 6, 166--179 (2024).

\bibitem{Tong2003}
Tong, L.~M. et al. Subwavelength-diameter silica wires for low-loss optical wave guiding.
\emph{Nature} 426, 816--819 (2003).

\bibitem{Sague2007}
Sagu\'{e}. et al. Cold-Atom Physics Using Ultrathin Optical Fibers: Light-Induced Dipole Forces and Surface Interactions.
\emph{Physical Review Letters} 99, 163602 (2007).

\bibitem{Brambilla2010}
Brambilla, G. et al. Optical fibre nanowires and microwires: a review.
\emph{J. Opt. A} 12, 043001 (2010).

\bibitem{Tong2012}
Tong, L.~M. et al. Optical microfibers and nanofibers: A tutorial.
\emph{Optics Communications} 285, 4641--4647 (2012).

\bibitem{Solano2017a}
Solano, P. et al. Optical Nanofibers: A New Platform for Quantum Optics.
\emph{Advances in Atomic, Molecular and Optical Physics} 66, 439--505 (2017).

\bibitem{Nayak2018}
Nayak, K.~P. et al. Nanofiber quantum photonics.
\emph{Journal of Optics} 20, 073001 (2018).

\bibitem{Li2024}
Li, W. et al. Atom-light interactions using optical nanofibres--a perspective.
\emph{J. Phys. Photonics} 6, 021002 (2024).

\bibitem{Love1986}
Love, J.~D. et al. Quantifying loss minimisation in single-mode fibre tapers.
\emph{Electron. Lett.} 22, 912--914 (1986).

\bibitem{Love1991}
Love, J.~D. et al. Tapered single-mode fibres and devices. Part 1: Adiabaticity criteria.
\emph{IEE Proc.} 138, 343--354 (1991).

\bibitem{Birks1992}
Birks T.~A. et al. The shape of fiber tapers.
\emph{J. Lightwave Technol.} 10, 432--438 (1992).

\bibitem{Hoffman2014}
Hoffman J.~E. et al. Ultrahigh transmission optical nanofibers.
\emph{AIP Adv.} 4, 067124 (2014).

\bibitem{Nagai2014}
Nagai R. et al. Ultra-low-loss tapered optical fibers with minimal lengths.
\emph{Opt. Express} 22, 28427--28436 (2014).

\bibitem{Ruddell2020}
Ruddell, S.~K. et al. Ultra-low-loss nanofiber Fabry-Perot cavities optimized for cavity quantum electrodynamics.
\emph{Optics Letters} 45, 4875 (2020).

\bibitem{Kien2004}
Kien, F.~L. et al. Atom trap and waveguide using a two-color evanescent field around a subwavelength-diameter optical fiber..
\emph{Physical Review A} 70, 063403 (2004).

\bibitem{Vetsch2010}
Vetsch, E. et al. Optical interface created by laser-cooled atoms trapped in the evanescent field surrounding an optical nanofiber.
\emph{Physical Review Letters} 104, 203603 (2010).

\bibitem{Goban2012}
Goban, A. et al. Demonstration of a state-insensitive, compensated nanofiber trap.
\emph{Physical Review Letters} 109, 033603 (2012).

\bibitem{Kestler2023}
Kestler, G. et al. State-insensitive trapping of alkaline-earth atoms in a nanofiber-based optical dipole trap.
\emph{PRX Quantum} 4, 040308 (2023).

\bibitem{Mitsch2014}
Mitsch, R. et al. Quantum state-controlled directional spontaneous emission of photons into a nanophotonic waveguide.
\emph{Nature Communications} 5, 5713 (2014).

\bibitem{Sayrin2015}
Sayrin, C.. et al Storage of fiber-guided light in a nanofiber-trapped ensemble of cold atoms.
\emph{Optica} 2, 353--356 (2015).

\bibitem{Gouraud2015}
Gouraud, B. et al. Demonstration of a Memory for Tightly Guided Light in an Optical Nanofiber.
\emph{Physical Review Letters} 114, 180503 (2015).

\bibitem{Corzo2019}
Corzo, N.~V. et al. Waveguide-coupled single collective excitation of atomic arrays.
\emph{Nature} 566, 359--362 (2019).

\bibitem{Liedl2023}
Liedl, C. et al. Collective Radiation of a Cascaded Quantum System: From Timed Dicke States to Inverted Ensembles.
\emph{Physical Review Letters} 130, 163602 (2023).

\bibitem{Endres2016}
Endres, M. et al. Atom-by-atom assembly of defect-free one-dimensional cold atom arrays.
\emph{Science} 354, 1024--1027 (2016).

\bibitem{Barredo2016}
Barredo, D. et al. An atom-by-atom assembler of defect-free arbitrary two-dimensional atomic arrays.
\emph{Science} 354, 1021--1023 (2016).

\bibitem{Manetsch2025}
Manetsch, H.~J. et al. A tweezer array with 6,100 highly coherent atomic qubits.
\emph{Nature} 647, 60--67 (2025).

\bibitem{Kusano2025}
Kusano, T. et al. Plane-selective manipulations of nuclear spin qubits in a three-dimensional optical tweezer array.
\emph{Physical Review Research} 7, L022045 (2025).

\bibitem{Ebadi2022}
Ebadi, S. et al. Quantum optimization of maximum independent set using Rydberg atom arrays.
\emph{Science} 376, 1209--1215 (2022).

\bibitem{Bluvstein2022}
Bluvstein, D. et al. A quantum processor based on coherent transport of entangled atom arrays.
\emph{Nature} 604, 451--456 (2022).

\bibitem{Graham2022}
Graham, T.~M. et al. Multi-qubit entanglement and algorithms on a neutral-atom quantum computer.
\emph{Nature} 604, 457--462 (2022).

\bibitem{Singh2023}
Singh, K. et al. Mid-circuit correction of correlated phase errors using an array of spectator qubits.
\emph{Science} 380, 1265--1269 (2023).

\bibitem{Evered2023}
Evered, S.~J. et al. High-fidelity parallel entangling gates on a neutral-atom quantum computer.
\emph{Nature} 622, 268--272 (2023).

\bibitem{Scholl2023}
Scholl, P. et al. Erasure conversion in a high-fidelity Rydberg quantum simulator.
\emph{Nature} 622, 273--278 (2023).

\bibitem{Ma2023}
Ma, S. et al. High-fidelity gates and mid-circuit erasure conversion in an atomic qubit.
\emph{Nature} 622, 279--284 (2023).

\bibitem{Bluvstein2024}
Bluvstein, D. et al. Logical quantum processor based on reconfigurable atom arrays.
\emph{Nature} 626, 58--65 (2024).

\bibitem{Cao2024}
Cao, A. et al. Multi-qubit gates and Schr\"odinger cat states in an optical clock.
\emph{Nature} 634, 315--320 (2024).

\bibitem{Finkelstein2024}
Finkelstein, R. et al. Universal quantum operations and ancilla-based read-out for tweezer clocks.
\emph{Nature} 634, 321--327 (2024).

\bibitem{Grinkemeyer2025}
Grinkemeyer, B. et al. Error-detected quantum operations with neutral atoms mediated by an optical cavity.
\emph{Science} 387, 1301--1305 (2025).

\bibitem{Evered2025}
Evered, S.~J. et al. Probing the Kitaev honeycomb model on a neutral-atom quantum computer.
\emph{Nature} 645, 341--347 (2025).

\bibitem{Chiu2025}
Chiu, N.-C. et al. Continuous operation of a coherent 3,000-qubit system.
\emph{Nature} 646, 1075--1080 (2025).

\bibitem{Bluvstein2026}
Bluvstein, D. et al. A fault-tolerant neutral-atom architecture for universal quantum computation.
\emph{Nature} 649, 39--46 (2026).

\bibitem{Browaeys2020}
Browaeys, A. et al. Many-body physics with individually controlled rydberg atoms.
\emph{Nature Physics} 16, 132--142 (2020).

\bibitem{Ebadi2021}
Ebadi, S. et al. Quantum phases of matter on a 256-atom programmable quantum simulator.
\emph{Nature} 595, 227--232 (2021).

\bibitem{Scholl2021}
Scholl, P. et al. Quantum simulation of 2D antiferromagnets with hundreds of Rydberg atoms.
\emph{Nature} 595, 233--238 (2021).

\bibitem{Spar2022}
Spar, B.~M. et al. Realization of a fermi-hubbard optical tweezer array.
\emph{Physical Review Letters} 128, 223202 (2022).

\bibitem{Menon2024}
Menon, S.~G. et al. An integrated atom array-nanophotonic chip platform with background-free imaging.
\emph{Nature Communications} 15, 6156 (2024).

\bibitem{Djordjevic2021}
Djordjevic, T. et al. Entanglement transport and a nanophotonic interface for atoms in optical tweezers.
\emph{Science} 373, 1511--1514 (2021).

\bibitem{Sunami2025}
Sunami, S. et al. Scalable networking of neutral-atom qubits: Nanofiber-based approach for multiprocessor fault-tolerant quantum computers.
\emph{PRX Quantum} 6, 010101 (2025).

\bibitem{Sinclair2025}
Sinclair, J. et al. Fault-tolerant optical interconnects for neutral-atom arrays.
\emph{Physical Review Research} 7, 013313 (2025).

\bibitem{Main2025}
Main, D. et al. Distributed quantum computing across an optical network link.
\emph{Nature} 638, 383--388 (2025).

\bibitem{Ji2026}
Ji, J.~-W. et al. Global quantum network with ground-based single-atom memories in optical cavities and satellite links.
\emph{Physical Review Applied} 25, 024050 (2026).

\bibitem{Chang2012}
Chang, D.~E. et al. Cavity QED with atomic mirrors.
\emph{New Journal of Physics} 14, 063003 (2012).

\bibitem{Chang2013}
Chang, D.~E. et al. Self-organization of atoms along a nanophotonic waveguide.
\emph{Physical Review Letters} 110, 113606 (2013).

\bibitem{Douglas2015}
Douglas, J.~S. et al. Quantum many-body models with cold atoms coupled to photonic crystals.
\emph{Nature Photonics} 9, 326 (2015).

\bibitem{Gonzalez-Tudela2015}
Gonz\'{a}lez-Tudela, A. et al. Deterministic generation of arbitrary photonic states assisted by dissipation.
\emph{Physical Review Letters} 115, 163603 (2015).

\bibitem{Corzo2016}
Corzo, N.~V. et al. Large Bragg reflection from one-dimensional chains of trapped atoms near a nanoscale waveguide.
\emph{Physical Review Letters} 117, 133603 (2016).

\bibitem{Lodahl2017}
Lodahl, P. et al. Chiral quantum optics.
\emph{Nature} 541, 473--480 (2017).

\bibitem{Solano2017b}
Solano, P. et al. Super-radiance reveals infinite-range dipole interactions through a nanofiber.
\emph{Nature Communications} 8, 1857 (2017).

\bibitem{Bello2019}
Bello, M. et al. Unconventional quantum optics in topological waveguide QED.
\emph{Sci. Adv.} 5, eaaw0297 (2019).

\bibitem{Albrecht2019}
Albrecht, A. et al. Subradiant states of quantum bits coupled to a one-dimensional waveguide.
\emph{New Journal of Physics} 21, 025003 (2019).

\bibitem{Mahmoodian2020}
Mahmoodian, S. et al. Dynamics of many-body photon bound states in chiral waveguide QED.
\emph{Phys. Rev. X} 10, 031011 (2020).

\bibitem{Fayard2021}
Fayard, N. et al. Many-body localization in waveguide QED.
\emph{Phys. Rev. Res.} 3, 033233 (2021).

\bibitem{Reicherter1999}
Reicherter, M. et al. Optical particle trapping with computer-generated holograms written on a liquid-crystal display.
\emph{Optics Letters}, 24, 608--610 (1999).

\bibitem{McKeever2003}
McKeever, J. et al. State-Insensitive Cooling and Trapping of Single Atoms in an Optical Cavity.
\emph{Physical Review Letters} 90, 133602 (2003).

\bibitem{Thompson2013}
Thompson, J.~D. et al. Coupling a single trapped atom to a nanoscale optical cavity.
\emph{Science} 340, 1202--1205 (2013).

\bibitem{Nayak2019}
Nayak, K.~P. et al. Real-time observation of single atoms trapped and interfaced to a nanofiber cavity.
\emph{Physical Review Letters} 123, 213602 (2019).

\bibitem{Kien2005}
Kien, F.~L. et al. Spontaneous emission of a cesium atom near a nanofiber: Efficient coupling of light to guided modes.
\emph{Physical Review A} 72, 032509 (2005).

\bibitem{Miki2013}
Miki, S. et al. High performance fiber-coupled nbtin superconducting nanowire single photon detectors with gifford-mcmahon cryocooler.
\emph{Optics Express} 21, 10208--10214 (2013).

\bibitem{Kimble1977}
Kimble, H.~J. et al. Photon antibunching in resonance fluorescence.
\emph{Physical Review Letters} 39, 691--695 (1977).

\bibitem{Short1983}
Short, R.. et al Observation of Sub-Poissonian Photon Statistics.
\emph{Physical Review Letters} 51, 384--387 (1983).

\bibitem{Schlosser2001}
Schlosser, N. et al. Sub-poissonian loading of single atoms in a microscopic dipole trap.
\emph{Nature} 411, 1024--1027 (2001).

\bibitem{Meng2018}
Meng, Y. et al. Near-ground-state cooling of atoms optically trapped 300^^c2^^a0nm away from a hot surface.
\emph{Physical Review X} 8, 031054 (2018).

\bibitem{Dicke1954}
Dicke, R.~H. Coherence in spontaneous radiation processes.
\emph{Physical Review} 93, 99--110 (1954).

\bibitem{Leonardo2007}
Di Leonardo, R. et al. Computer generation of optimal holograms for optical trap arrays.
\emph{Optics Express} 15, 1913--1922 (2007).

\bibitem{Derevianko1999}
Derevianko, A. et al. High-Precision Calculations of Dispersion Coefficients, Static Dipole Polarizabilities, and Atom-Wall Interaction Constants for Alkali-Metal Atoms.
\emph{Phys. Rev. Lett.} 82, 3589 (1999).

\bibitem{WallsMilburn}
Walls, D.~F. and Milburn, G.~J. ``Quantum Optics'', Springer.

\end{thebibliography}

\end{document}